\documentclass[11pt]{article}%
\usepackage{amsmath}
\usepackage{amssymb}
\usepackage{amsfonts}

\usepackage{cite}
\usepackage{graphicx}
\usepackage{float}
\usepackage{longtable}
\usepackage[utf8]{inputenc}
\usepackage{hyperref}

\usepackage{algorithmicx}
\usepackage[ruled,vlined]{algorithm2e}
\SetAlgoCaptionSeparator{ }

\usepackage{fullpage}
\usepackage{authblk}

\newcommand{\fref}[1]{Fig.~\ref{#1}}

\newcommand{\tref}[1]{Table~\ref{#1}}
\newcommand{\sref}[1]{Section~\ref{#1}}

\newenvironment{algo}[1][!h]
  {
   \begin{algorithm}[#1]%
  }{\end{algorithm}}

\newenvironment{method}[1][!h]
  {
   \begin{algorithm}[#1]%
  }{\end{algorithm}}

\providecommand{\U}[1]{\protect\rule{.1in}{.1in}}
\newtheorem{problem}{Problem}

\begin{document}

\title{Topology Adaption for the Quantum Internet}
\author[1,2,3]{Laszlo Gyongyosi\footnote{Email: \href{mailto:l.gyongyosi@soton.ac.uk}{l.gyongyosi@soton.ac.uk}. Parts of this work were presented in conference proceedings [9].}}
\author[2]{Sandor Imre}
\affil[1]{School of Electronics and Computer Science, University of Southampton, Southampton, SO17 1BJ, UK}
\affil[2]{Department of Networked Systems and Services, Budapest University of Technology and Economics, Budapest, H-1117 Hungary}
\affil[3]{MTA-BME Information Systems Research Group, Hungarian Academy of Sciences, Budapest, H-1051 Hungary}
\date{}

\maketitle
\begin{abstract}
In the quantum repeater networks of the quantum Internet, the varying stability of entangled quantum links makes dynamic topology adaption an emerging issue. Here we define an efficient topology adaption method for quantum repeater networks. The model assumes the random failures of entangled links and several parallel demands from legal users. The shortest path defines a set of entangled links for which the probability of stability is above a critical threshold. The scheme is utilized in a base-graph of the overlay quantum network to provide an efficient shortest path selection for the demands of all users of the network. We study the problem of entanglement assignment in a quantum repeater network, prove its computational complexity, and show an optimization procedure. The results are particularly convenient for future quantum networking, quantum-Internet, and experimental long-distance quantum communications.
\end{abstract}

\section{Introduction}
In the quantum Internet \cite{ref1,ref2,ref3,ref4}, noise, eavesdropping and other random failures on the entangled links lead to a time-varying link stability, which makes topology adaption of quantum repeater networks an emerging issue \cite{ref1,ref2,ref3,ref4,ref5,ref6,ref7,ref8,ref9,ref10,ref11,ref12, ref16,ref17,ref18,ref19,ref20,ref21,ref22,ref23,ref24,ref25,ref26,ref27,ref28,ref29,ref30}. Because the failures of the entangled links are random, the actual shortest paths must be updated dynamically, which is particularly important for a seamless transmission of quantum information. In a quantum Internet scenario, the quantum repeater nodes are connected through several different entanglement levels \cite{ref1,ref2,ref3,ref4,ref8,ref9,ref10,ref11,ref12,ref20,ref21,ref22,ref23,ref24,ref30,ref31,ref32,ref33,ref34,ref35,ref36,ref37,ref38,ref39,ref40,ref41,ref42,ref43,ref44,ref45} making the problem harder. In this multi-level quantum network, the entanglement level of a link between the quantum nodes refers to the level of entanglement shared between the source and the target nodes incident to that link. The level of an entangled link not only identifies the number of spanned quantum nodes (hop-distance), but also provides a base-point on the stability of that link between the nodes. 

Here, we define a dynamic topology adapting method to efficiently manage the random link errors of the quantum links of the quantum Internet. The vulnerability of the entangled links requires topology adaption to provide seamless communication between the quantum nodes. The quantum links are characterized by a given probability of existence (stability) in an overlay quantum network, which describes the probability of a given level of entanglement shared between nodes. The proposed adapting algorithm determines a set of entangled links in a proactive way for the formulation of a new shortest path in the updated topology of the overlay entangled quantum network. Specifically, to minimize the risk of a possible failure, we introduce a threshold parameter for each entanglement level to construct alternative paths in the network before a given entangled link becomes critical. In particular, the shortest path is determined in a base-graph, which allows us to perform efficient decentralized routing in the quantum repeater network \cite{ref8}. The base-graph is a $k$-dimensional $n$-size lattice graph that contains the mapped entangled overlay quantum network so that all nodes and connections are preserved. The base-graph allows us to model the entangled link in a distance-based manner and to perform efficient decentralized routing in the network.  

The routing metric in the base-graph is the number of links between a source and the target (diameter), while the set of new entangled links is determined with respect to their probability of existence. Because the probability of the stability of a given quantum link depends on the level of entanglement shared through that link, the metric of dynamic adaption depends on the actual overlay quantum network setting. Therefore, the algorithm provides a consecutive update of quantum links and prepares a network for changing link conditions to improve the reliability. 

The updated configuration of links provides an input for the decentralized routing scheme, which is performed purely in a base-graph \cite{ref8}. We assume a multiuser scenario with multiple parallel demands in the network; therefore, the task of the dynamic topology adaption algorithm is to maximize the aggregate link probability over all the demanded paths of the users. 

We also study the problem of entanglement assignment and define an optimization procedure. The problem with entanglement assignment is that an intermediate quantum repeater node occurs if the demands of two or more source nodes with respect to a given target node are interfering; i.e., several source nodes would like to reserve a given resource entangled state in an intermediate quantum node to establish entanglement with a target node through the intermediate node by the operation of entanglement swapping. The solution to this problem requires the definition of disjoint sets in an intermediate quantum node, a set of entangled resource states (resource set), and a set of interfering entangled states (interference set). With the utilization of our base-graph approach, we characterize the computational complexity of the problem and prove that efficient optimized solutions exist. 

This paper is organized as follows. In \sref{sec2}, we discuss the proposed topology adaption algorithm for the quantum Internet. In \sref{sec3}, we study the problem of entanglement assignment and provide an optimization for managing interfering demands. Finally, \sref{sec4} concludes the results. Supplemental information is included in the Appendix.

\section{Topology Adaption}
\label{sec2}
Let $N$ be an entangled overlay quantum network, $N=(V,{\rm {\mathcal S}})$, where ${\rm {\mathcal S}}$ is an initial set of entangled links ${\rm {\mathcal S}}=\{E_{i} \}$, $i=0,\ldots ,h-1$, where $E_{i} $ is ${\rm L}_{l} $-level entangled link, $l=1,2,\ldots ,r$, while $V$ is the set of quantum nodes of the overlay quantum network $N$. Assuming a quantum repeater network with the doubling architecture \cite{ref1,ref2,ref3,ref4}, for an ${\rm L}_{l} $-level entangled link the $d(x,y)_{{\rm L}_{l} } $ hop distance between $x$ and $y$, where $x,y\in V$, is $d(x,y)_{{\rm L}_{l} } =2^{l-1} $ \cite{ref1,ref8}, and each ${\rm L}_{l} $-level entangled link $E(x,y)$ can be established only with a given probability, ${{\Pr }_{{{\text{L}}_{l}}}}( E( x,y ) )$, that is defined without loss of generality as
\begin{equation} \label{1)} 
{{\Pr }_{{{\text{L}}_{l}}}}( E( x,y ) )=\Pr (S(E(x,y)))\cdot (1-\Pr ({\rm {\mathcal L}}(E(x,y))))\cdot F({| \psi ^{(E(x,y))} \rangle}),             
\end{equation} 
where $\Pr (S(E(x,y)))$ is the success probability of the $S(E(x,y))$ entanglement swapping procedure between $x$ and $y$, $\Pr ({\rm {\mathcal L}}(E(x,y)))$ is the photon loss probability of the quantum link between nodes $x$ and $y$, while $F({| \psi ^{(E(x,y))}  \rangle})$ is the fidelity of the $d$-dimensional maximally entangled system ${| \psi ^{(E(x,y))} \rangle} $ associated to link $E(x,y)$, respectively. 

Note, our proposed scheme can be extended to next-generation quantum repeater networks \cite{ref34} that do not rely on the doubling architecture \cite{ref1,ref8}. 

The aim of the link configuration algorithm is to determine the shortest path between a transmitter node $A$ and a target node $B$ via the updated link set ${\rm {\mathcal S}}^{{\rm *}} =\{E_{i}^{{\rm *}} \}$, $i=0,\ldots ,h^{{\rm *}} -1$ of ${\rm L}_{l} $-level entangled links, where set ${\rm {\mathcal S}}^{{\rm *}} $ identifies those links which have the highest probability in the network. Each link of ${\rm {\mathcal S}}^{{\rm *}} $ is characterized by a metric called entanglement throughput, which specifically is the number of $d$-dimensional maximally entangled states per second of a particular fidelity $F$ \cite{ref1}, and it is denoted by $Q^{(F)} (E_{i}^{{\rm *}} )$ for a given $E_{i}^{{\rm *}} $ in the modified entangled overlay quantum network $N=(V,{\rm {\mathcal S}}^{{\rm *}} )$.

To describe the entangled connections between the quantum nodes, we characterize a base-graph. The base-graph is an abstracted graph that contains all information about the overlay (physical) quantum network (node positions, hop-distances, established entangled links, link properties). As it has been shown in \cite{ref8}, the base-graph approach allows us to perform efficient decentralized routing in a quantum Internet scenario.

Let $G^{k} $ be a $k$-dimensional $n$-size base-graph of an entangled overlay quantum network $N$ \cite{ref8}. Let $x,y\in V$ two nodes in $N$, and let $\phi (x)$ and $\phi (y)$ be the maps of $x,y$ in $G^{k} $, where $\phi :V\to G^{k} $ is a mapping function \cite{ref8} that maps from $V$ onto $G^{k} $. As it can be verified \cite{ref8}, the $p(\phi (x),\phi (y))$ probability that $\phi (x)$ and $\phi (y)$ are connected through an ${\rm L}_{l} $-level entanglement in $G^{k} $ is as 
\begin{equation} \label{ZEqnNum204871} 
p(\phi (x),\phi (y))=\frac{d(\phi (x),\phi (y))^{-k} }{H_{n} } +c_{\phi (x),\phi (y)} , 
\end{equation} 
where $d(\phi (x),\phi (y))$ is the L1 distance between $\phi (x)$ and $\phi (y)$ in $G^{k} $, $H_{n} =\sum _{z}d(\phi (x),\phi (z)) $ is a normalizing term \cite{ref8} taken over all entangled contacts of node $\phi (x)$ in $G^{k} $, while $c_{\phi (x),\phi (y)} $ is a constant defined \cite{ref8,ref13,ref14} as
\begin{equation} \label{3)} 
c_{\phi (x),\phi (y)} ={{\Pr }_{{{\text{L}}_{l}}}}( E( x,y ) )-\frac{d(\phi (x),\phi (y))^{-k} }{H_{n} } , 
\end{equation} 
where ${{\Pr }_{{{\text{L}}_{l}}}}( E( x,y ) )$ is the probability that nodes $x,y\in V$ are connected through an ${\rm L}_{l} $-level entangled link $E(x,y)$ in the overlay quantum network $N$. 

Due to noise or other random link failure, the ${{\Pr }_{{{\text{L}}_{l}}}}( E( x,y ) )$ probability of a given ${\rm L}_{l} $-level entangled link $E(x,y)$ is not constant in the overlay quantum network $N$. This fact has the consequence that an actual shortest path determined for set ${\rm {\mathcal S}}$ cannot be used further, and a new shortest path needs to be determined. The aim of the adaption algorithm is to update the $p(\phi (x),\phi (y))$ probabilities in the base-graph according to a threshold $\Pr _{{\rm L}_{l} }^{{\rm *}} $ for a given  ${\rm L}_{l} $-level entangled link $E(x,y)$ to determine the actual shortest path for ${\rm {\mathcal S}}^{{\rm *}} $. Only that $E(x,y)$ link will be included in the updated set ${\rm {\mathcal S}}^{{\rm *}} $ of links for which the condition $\Pr _{{\rm L}_{l} } (x,y)\ge \Pr _{{\rm L}_{l} }^{{\rm *}} $ holds. The algorithm is performed for all entangled contacts of a given node $\phi (x)$ for $\forall x$. Then, we apply the decentralized routing method ${\rm {\mathcal A}}$ defined in \cite{ref8} to determine the shortest path by at most ${\rm {\mathcal O}}(\log n)^{2} $ steps in a $k$-dimensional $n$-size base-graph $G^{k} $.

The topology adaption algorithm ${\rm {\mathcal R}}$ in $G^{k} $ is given in Algorithm 1. 

 \setcounter{algocf}{0}
\begin{algo}
  \DontPrintSemicolon
\caption{\textit{Topology Adaption.}}
\textbf{Step 1}. Let $\phi (A)$ refer to the map of a source node $A$ in the overlay network $N$, and let $\phi (B)$ be the map of a target node $B\in V$ in $G^{k} $. For each pair of neighboring nodes $\phi (x)$ and $\phi (y)$ in $G^{k} $, estimate ${{\Pr }_{{{\text{L}}_{l}}}}( E( x,y ) )$ of an ${\rm L}_{l} $-level entangled link $E(x,y)$, where $x$ and $y$ are nodes in the overlay network $N$, while ${\rm L}_{l} $ is the level of entanglement between $x$ and $y$. Let ${\rm {\mathcal S}}$ refer to the initial set of entangled links. 

\textbf{Step 2}. Using a threshold probability $\Pr _{{\rm L}_{l} }^{{\rm *}} $ of an ${\rm L}_{l} $-level entangled link, update $p(\phi (x),\phi (y))$ for all node pairs $\{\phi (x),\phi (y)\}\in G^{k} $ via the following rule: 
\[p\left(\phi \left(x\right),\phi \left(y\right)\right)^{{\rm *}} =\left\{\begin{array}{l} {\frac{d\left(\phi \left(x\right),\phi \left(y\right)\right)^{-k} }{H_{n} } +c_{\phi \left(x\right),\phi \left(y\right)}^{{\rm *}} ,{\rm \; }\text{if}{\rm \; }\Pr _{{\rm L}_{l} } \left(x,y\right)\ge \Pr _{{\rm L}_{l} }^{{\rm *}} } \\ {0,{\rm \; otherwise\; \; \; \; \; \; \; \; \; \; \; \; \; \; \; \; \; \; \; \; \; \; \; \; \; \; \; \; \; \; \; \; \; \; \; \; \; \; \; \; \; \; \; \; \; \; \; \; \; \; }} \end{array}\right. ,\] 
where 
$c_{\phi \left(x\right),\phi \left(y\right)}^{{\rm *}} =\Pr _{{\rm L}_{l} }^{{\rm *}} \left(E\left(x,y\right)\right)-\frac{d\left(\phi \left(x\right),\phi \left(y\right)\right)^{-k} }{H_{n} }.$ 

\textbf{Step 3}. Using the results of Step 2, determine a new set ${\rm {\mathcal S}}^{{\rm *}} $ of entangled links for all node pairs of $G^{k} $.

\textbf{Step 4}. Using the new link configuration ${\rm {\mathcal S}}^{{\rm *}} $ determined in Step 3, find the new shortest path via ${\rm {\mathcal A}}$ of \cite{ref8} between source node $\phi (A)$ and target node $\phi (B)$ in terms of diameter $D(\phi (A),\phi (B))$.

\end{algo}

\begin{center}
\begin{figure*}[!h]
%\vspace{-0.5cm}
\begin{center}
\includegraphics[angle = 0,width=1\linewidth]{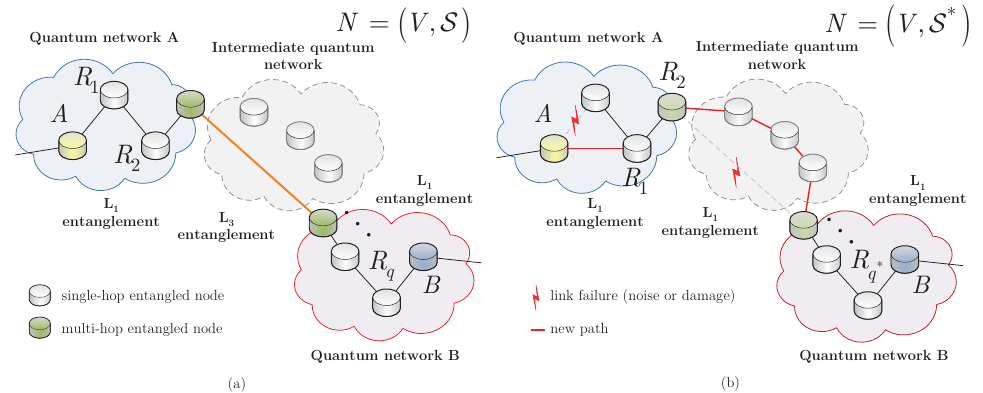}
\caption{The topology adaption scheme in an overlay quantum repeater network. The initial network $N=(V,{\rm {\mathcal S}})$ consists of a transmitter node (A) a distant target node (B), and $q$ intermediate repeater nodes $R_{i} ,i=1,\ldots ,q$. The set of ${\rm {\mathcal S}}$ contains ${\rm L}_{l} $-level entangled links, $l=1,2,\ldots ,r$. Nodes $A$ and $B$ have an actual shortest path in ${\rm {\mathcal S}}$ (a). The random link failures (noise, other link damage) leads to modified overlay network $N=(V,{\rm {\mathcal S}}^{{\rm *}} )$, with $R_{i} ,i=1,\ldots ,q^{{\rm *}} $ repeater nodes and link set ${\rm {\mathcal S}}^{{\rm *}} $. The new shortest path (newer links are depicted by the red lines) is determined by decentralized routing in the base-graph $G^{k} $ (b).} 
 \label{fig1}
 \end{center}
\end{figure*}
\end{center}

The ${\rm {\mathcal R}}$ topology adaption algorithm performed on an initial overlay quantum network $N=(V,{\rm {\mathcal S}})$ is illustrated in \fref{fig1}. The algorithm ${\rm {\mathcal S}}^{{\rm *}} $ in the $k$-dimensional $n$-size base-graph $G^{k} $ results in the modified topology overlay network $N=(V,{\rm {\mathcal S}}^{{\rm *}} )$. A shortest path for ${\rm {\mathcal S}}^{{\rm *}} $ is determined by the decentralized routing algorithm ${\rm {\mathcal A}}$ of \cite{ref8}.

\section{Entanglement Assignment}
\label{sec3}
In this section, we formulate the problem of entanglement assignment for an entangled overlay quantum network $N=(V,{\rm {\mathcal S}}^{{\rm *}} )$, where ${\rm {\mathcal S}}^{{\rm *}} $ is determined by the ${\rm {\mathcal R}}$ topology adaption algorithm in the $G^{k} $, the $k$-dimensional base-graph of the entangled overlay quantum network $N$. 

Let $\phi (A)_{U_{k} } $ and $\phi (B)_{U_{k} } $ refer to the source and target nodes associated with the demand of a user $U_{k} $ in $G^{k} $, $k=1\ldots K$. Let $E_{h} $ refer to the entangled link $E_{h} (\phi (x_{i} ),\phi (x_{j} ))$, and let ${\rm {\mathcal L}}_{h} $ be the set (resource set) of $d$-dimensional maximally entangled states shared between $\phi (x_{i} ),\phi (x_{j} )$ through link $E_{h} $, as \cite{ref1, ref8}
\begin{equation} \label{4)} 
{\rm {\mathcal L}}_{h} =\{{| \psi _{0}  \rangle} ,\ldots ,{| \psi _{g-1}  \rangle} \},                                                
\end{equation} 
where $g$ is the number of entangled states established between nodes $\phi (x_{i} )$ and $\phi (x_{j} )$ through $E_{h} $. 

Let the set ${\rm {\mathcal S}}^{{\rm *}} =\{E_{h}^{{\rm *}} \},$$h=0,\ldots ,h^{{\rm *}} -1$ of entangled links determined in $G^{k} $ by Step 2 of ${\rm {\mathcal R}}$. The problem of the entanglement assignment between $\phi (A)_{U_{k} } $ and $\phi (B)_{U_{k} } $ for all $U_{k} $ is subject to the following minimization, precisely:
\begin{equation} \label{ZEqnNum542690} 
\begin{split}
   \zeta ( C )&=\min \sum\limits_{k\in K}{\sum\limits_{f\in {{\mathcal{L}}_{h}}}{\sum\limits_{h\in E}{( 1-{{\Pr }_{{{\text{L}}_{l}}}}( E_{h}^{\text{*}} ) )C_{k,h}^{f}}}} \\ 
  &=\min \sum\limits_{k\in K}{\sum\limits_{f\in {{\mathcal{L}}_{h}}}{\sum\limits_{h\in E}{( 1-p_{h}^{\text{*}} )C_{k,h}^{f}}}},  
\end{split}
\end{equation} 
where $K$ is the number of users, ${\rm {\mathcal L}}_{h} $ is the resource set of $d$-dimensional maximally entangled states available through the entangled link $E_{h}^{{\rm *}} \in {\rm {\mathcal S}}^{{\rm *}} $, $E$ is a set of edges, $\Pr _{{\rm L}_{l} } (E_{h}^{{\rm *}} )$ is the probability of existence of the ${\rm L}_{l} $-level entangled link $E_{h}^{{\rm *}} $ in the overlay network $N$, $p_{h}^{{\rm *}} $ is the corresponding probability of the ${\rm L}_{l} $-level entangled link $E_{h}^{{\rm *}} $ in the base-graph $G^{k} $,  and $C_{k,h}^{f} $ is a variable which equals 1 if the $f\in {\rm {\mathcal L}}_{h} $ maximally entangled resource state ${| \psi _{f}  \rangle} $ is assigned to the $k$-th user $U_{k} $ through link $E_{h}^{{\rm *}} $ or 0 otherwise. 

Let $Q^{(F)} (E_{h}^{{\rm *}} )$ refer to the number of $d$-dimensional maximally entangled states per second of a particular fidelity F available through the entangled link $E_{h}^{{\rm *}} $, and let $Q^{(F)} (U_{k} )$ be the demand of user $U_{k} $ with respect to the number of $d$-dimensional maximally entangled states per second of a particular fidelity F through link $E_{h}^{{\rm *}} $. Specifically, for a given $E_{h}^{{\rm *}} $, the following relation holds \cite{ref15}:
\begin{equation} \label{6)} 
\sum _{k\in K}\sum _{f\in {\rm {\mathcal L}}_{h} }C_{k,h}^{f} Q^{(F)} (U_{k} )\le Q^{(F)} (E_{h}^{{\rm *}} )  {\kern 1pt} {\kern 1pt} . 
\end{equation} 
As it can be verified via the flow conservation rules (Kirchhoff's law) \cite{ref15}, for all users $U_{k} $, a quantity $\Delta (C_{k,h}^{f} )$ can be defined with respect to $C_{k,h}^{f} $, without loss of generality, as
\begin{equation} \label{7)} 
\Delta (C_{k,h}^{f} )=\sum _{f\in {\rm {\mathcal L}}_{h} }\sum _{{\rm {\mathcal W}}_{h,j} }C_{k,h}^{f}   -\sum _{f\in {\rm {\mathcal L}}_{h} }\sum _{{\rm {\mathcal W}}_{h,i} }C_{k,h}^{f}   , 
\end{equation} 
where
\begin{equation} \label{ZEqnNum413381} 
\begin{split}
   {{\mathcal{W}}_{h,j}}:&h\in \{ h:E_{h}^{\text{*}}( \phi {{( {{x}_{t}} )}_{{{U}_{k}}}},\phi {{( {{x}_{j}} )}_{{{U}_{k}}}} )\in E;\\ 
 & \phi {{( {{x}_{j}} )}_{{{U}_{k}}}}\in V; \\ 
 & \phi {{( {{x}_{j}} )}_{{{U}_{k}}}}\ne \phi {{( {{x}_{t}} )}_{{{U}_{k}}}} \}, \\ 
\end{split}
\end{equation} 
where $\phi (x_{t} )_{U_{k} } \in G^{k} $ is a node associated to a given user $k\in K$, such that $E_{h}^{{\rm *}} (\phi (x_{t} )_{U_{k} } ,\phi (x_{j} )_{U_{k} } )$ is a link incident out of $\phi (x_{t} )_{U_{k} } \in G^{k} $, while $E_{h}^{{\rm *}} (\phi (x_{i} )_{U_{k} } ,\phi (x_{t} )_{U_{k} } )$ is a link incident onto node $\phi (x_{t} )_{U_{k} } \in G^{k} $ in the base-graph \cite{ref15}, while  
\begin{equation} \label{ZEqnNum969574} 
\begin{split}
   {{\mathcal{W}}_{h,i}}:&h\in \{ h:E_{h}^{\text{*}}( \phi {{( {{x}_{i}} )}_{{{U}_{k}}}},\phi {{( {{x}_{t}} )}_{{{U}_{k}}}} )\in E;\\ 
 & \phi {{( {{x}_{i}} )}_{{{U}_{k}}}}\in V; \\ 
 & \phi {{( {{x}_{i}} )}_{{{U}_{k}}}}\ne \phi {{( {{x}_{t}} )}_{{{U}_{k}}}} \}. \\ 
\end{split}
\end{equation} 
As one can readily check, using \eqref{ZEqnNum413381} and \eqref{ZEqnNum969574}, $\Delta (C_{k,h}^{f} )$ is yielded as 
\begin{equation} \label{10)} 
\Delta (C_{k,h}^{f} )=\left\{\begin{array}{l} {1,{\rm \; }\text{if}{\rm \; }\phi \left(x_{t} \right)_{U_{k} } =\phi \left(A\right)_{U_{k} } } \\ {-1,{\rm \; }\text{if}{\rm \; }\phi \left(x_{t} \right)_{U_{k} } =\phi \left(B\right)_{U_{k} } } \\ {0,{\rm \; otherwise\; \; \; \; \; \; \; \; \; \; \; \; \; \; \; \; }} \end{array}\right. , 
\end{equation} 
where $\phi (A)_{U_{k} } $ and $\phi (B)_{U_{k} } $ are the source and target nodes of $U_{k} $.

Let's assume the situation of a source node $\phi (A)_{U_{i} } =\phi (s_{i} )$ of user $U_{i} $, and a source node $\phi (B)_{U_{j} } =\phi (s_{j} )$ of user $U_{j} $ would like to share entanglement with a target node $\phi (y_{i} )$ through an $\phi (x_{i} )$ intermediate node. Let ${\rm {\mathcal L}}_{h} $ refer to the resource set of entangled states shared between $\phi (x_{i} )$ and $\phi (y_{i} )$ through an entangled link $E_{h}^{{\rm *}} $. Let each source node be associated to a given entangled state $f$ from ${\rm {\mathcal L}}_{h} $ (resource state) in $\phi (x_{i} )$.

Let's index the interfering queries of $\phi (s_{i} )$, $\phi (s_{j} )$ as $\{q,q'\}\in {\rm {\mathcal I}}$, where $q$ refers to the demand of $\phi (s_{i} )$, $q'$ is the interfering demand of $\phi (s_{j} )$, and ${\rm {\mathcal I}}$ is a set of interfering demands (interference set) with respect to $f\in {\rm {\mathcal L}}_{h} $. 

The next constraint assures that for a given maximally entangled resource state $f\in {\rm {\mathcal L}}_{h} $, at most one interfering entangled state (i.e., $q$ or $q'$) is assigned from ${\rm {\mathcal I}}$ for each $\{q,q'\}$ as
\begin{equation} \label{11)} 
\sum _{k\in K}K_{k,q}^{f}  +\sum _{k\in K}K_{k,q'}^{f}  \le 1,                                            
\end{equation} 
where $K_{k,h}^{f} $ is a variable equal to 1 if $q\in {\rm {\mathcal I}}$ is assigned to resource state $f\in {\rm {\mathcal L}}_{h} $ or 0 otherwise. The parallel serving of the interfering demands is trivial if $|{\rm {\mathcal I}}|\le |{\rm {\mathcal L}}_{h} |$. On the other hand, if $|{\rm {\mathcal I}}|>|{\rm {\mathcal L}}_{h} |$, the problem requires an optimization procedure to achieve an optimal assignment.   

\subsection{Optimization}

The optimal assignment of the interfering entangled states of ${\rm {\mathcal I}}$ at a given set of ${\rm {\mathcal S}}^{{\rm *}} =\{E_{h}^{{\rm *}} \},$$h=0,\ldots ,h^{{\rm *}} -1$ is characterized by the following constraints. 

The constraints of the optimization method is defined in Method 1.

 \setcounter{algocf}{0}
\begin{method}
  \DontPrintSemicolon
\caption{\textit{Optimization (assignment of interfering demands).}}
\begin{problem}
Determine ${\rm {\mathcal S}}^{{\rm *}} $ via ${\rm {\mathcal R}}$. At a given set ${\rm {\mathcal I}}$ and ${\rm {\mathcal L}}_{h} $, where $|{\rm {\mathcal I}}|>|{\rm {\mathcal L}}_{h} |$, determine the optimal assignment of the entangled states of ${\rm {\mathcal I}}$ in a given node $\phi (x_{i} )$. 
\end{problem}

\textbf{Constraint 1}. For a given resource state $f$ of set ${\rm {\mathcal L}}_{h} $, assign at most one entangled state $q$ from ${\rm {\mathcal I}}$. 

\textbf{Constraint 2}. If all resource states of ${\rm {\mathcal L}}_{h} $ are assigned, define multiple sets ${\rm {\mathcal L}}_{i} $, $0<i\le z-2$, where $z$ is the degree of $\phi (x_{i} )$. For a given resource state $w$ of ${\rm {\mathcal L}}_{i} $, assign at most one entangled state $u$ from ${\rm {\mathcal I}}$. Stop the procedure as $|{\rm {\mathcal L}}_{h} |+|{\rm {\mathcal L}}_{0} |+\ldots +|{\rm {\mathcal L}}_{i} |\ge |{\rm {\mathcal I}}|$. 

\textbf{Constraint 3}. To achieve the minimization of the number of resource states $|{\rm {\mathcal L}}_{h} |$ and $|{\rm {\mathcal L}}_{i} |$, $0<i\le z-1$ in $\phi (x_{i} )$, minimize $|{\rm {\mathcal I}}|$.

\end{method}

\begin{center}
\begin{figure*}[!h]
%\vspace{-0.5cm}
\begin{center}
\includegraphics[angle = 0,width=0.8\linewidth]{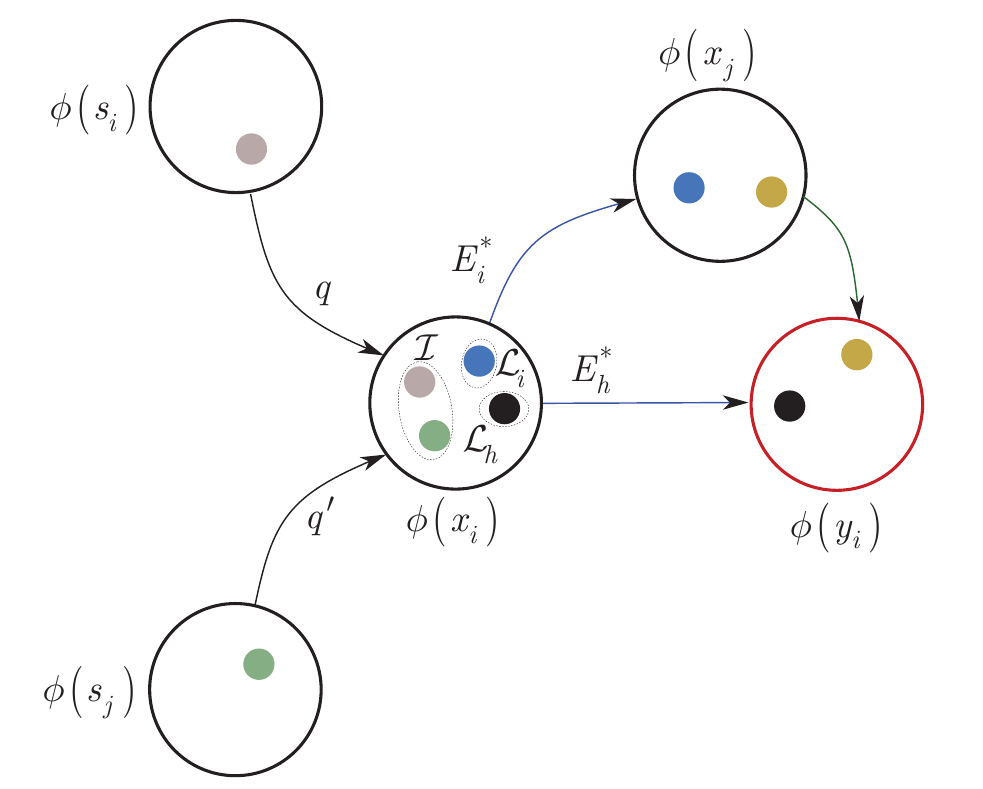}
\caption{A $|{\rm {\mathcal L}}_{h} |<|{\rm {\mathcal I}}|$ network situation in node $\phi (x_{i} )$ at a given ${\rm {\mathcal S}}^{{\rm *}} $. Source nodes $\phi (s_{i} )$, $\phi (s_{j} )$ are associated with the resource state ${| \psi _{f}  \rangle} $ from set ${\rm {\mathcal L}}_{h} =\{{| \psi _{f}  \rangle} \}$ via demands $\{q,q'\}\in {\rm {\mathcal I}}$ in an intermediate node $\phi (x_{i} )$ to establish entanglement with a distant target node $\phi (y_{i} )$. The conflicting nodes $\phi (s_{i} )$, $\phi (s_{j} )$ define a set of ${\rm {\mathcal I}}$ at a given resource set ${\rm {\mathcal L}}_{h} $.} 
 \label{fig2}
 \end{center}
\end{figure*}
\end{center}

The optimization problem for a quantum node $\phi (x_{i} )$ is illustrated in \fref{fig2}. Source nodes $\phi (s_{i} )$, $\phi (s_{j} )$ are associated with the resource entangled state ${| \psi _{f}  \rangle} $ from set ${\rm {\mathcal L}}_{h} =\{{| \psi _{f}  \rangle} \}$ via demands $\{q,q'\}\in {\rm {\mathcal I}}$ in an intermediate node $\phi (x_{i} )$ to establish entanglement with a distant target node $\phi (y_{i} )$. 

Set ${\rm {\mathcal L}}_{h} $ consists of the stored entangled resource states of $\phi (x_{i} )$, which are established between $\phi (x_{i} )$ and $\phi (y_{i} )$ through link $E_{h}^{{\rm *}} $, while set ${\rm {\mathcal L}}_{i} =\{{| \psi _{t}  \rangle} \}$ refers to the stored entangled resource states of $\phi (x_{i} )$ and $\phi (x_{j} )$ established via $E_{i}^{{\rm *}} $. The entangled resource states between $\{\phi (s_{i} ),\phi (x_{i} )\}$ and $\{\phi (s_{j} ),\phi (x_{i} )\}$ form the set ${\rm {\mathcal I}}$ of interfering entangled states. The link configuration ${\rm {\mathcal S}}^{{\rm *}} $ is determined by the link selection algorithm ${\rm {\mathcal R}}$. As the allocation of the resource entanglement is achieved, the entanglement is extended from the source nodes  $\phi (s_{i} )$, $\phi (s_{j} )$ to the target node $\phi (y_{i} )$ through the intermediate node $\phi (x_{i} )$ by the $U_{S} $ entanglement swapping operation (not shown). The $U_{S} $ operation is applied on the entangled states of ${\rm {\mathcal I}}$ and ${\rm {\mathcal L}}_{h} $ and ${\rm {\mathcal I}}$ and ${\rm {\mathcal L}}_{i} $, respectively.

\subsection{Computational Complexity}

Let's assume an entangled overlay quantum network $N$ with $|V|$ nodes. At a given interfering set ${\rm {\mathcal I}}$ and resource set ${\rm {\mathcal L}}_{h} $ of $E_{h}^{{\rm *}} $, the aim is to determine the optimal assignment of the entangled states of ${\rm {\mathcal I}}$. Let $|{\rm {\mathcal I}}|=k$, where $k$ is a minimized value. 

An assignment of the $k$ states of ${\rm {\mathcal I}}$ to a resource set ${\rm {\mathcal L}}_{h} $ of link $E_{h}^{{\rm *}} $ is referred as valid if ${\rm {\mathcal L}}_{h} $ consists of exactly $k$ entangled states, $|{\rm {\mathcal I}}|=|{\rm {\mathcal L}}_{h} |=k$. If $|{\rm {\mathcal L}}_{h} |<k$, then $|{\rm {\mathcal I}}|=k>|{\rm {\mathcal L}}_{h} |$, in which case the assignment is referred to as not valid. Assignments for all ${\rm {\mathcal L}}_{i} $ sets can be performed in at most ${\rm {\mathcal O}}(|E_{h}^{{\rm *}} |)\le {\rm {\mathcal O}}(|V|^{2} )$ steps by some fundamental theory. By a similar assumption, whether different states from ${\rm {\mathcal I}}$ are assigned to different states of given set ${\rm {\mathcal L}}_{i} $ can also be determined in at most ${\rm {\mathcal O}}(|V|^{2} )$ steps. In fact, these complexities prove that the problem of assignment of the entangled states belongs to the class of NP problems \cite{ref15}. 

In particular, it can also be verified that there exists a known NP-complete problem, which polynomially reduces to the entanglement assignment problem. Precisely, this problem is the optimal vertex-coloring of a graph of conflicts $G$ (graph-coloring) \cite{ref15} and is discussed as follows. 

\subsection{Polynomial Reduction}

Let $G=(V,E)$ a graph of conflicts, where $V$ is a set of vertices, while $E$ is a set of edges, $|E|=|{\rm {\mathcal S}}^{{\rm *}} |$. If two vertices $i,j\in V$ are connected by an edge $e(i,j)\in E$, the vertices $i,j$ are referred to as conflicting vertices. The problem is therefore to find the optimal assignment of colors to vertices from $V$ using exactly $k$ colors such that any two conflicting vertices of $V$ receive different colors. 

As it can be verified, identifying the vertices of $V$ by the entangled states of $\phi (x_{i} )$ via set ${\rm {\mathcal Z}}_{i} $, $i=0,\ldots ,|V|-1$,  where a set ${\rm {\mathcal Z}}_{i} $ consist of all entangled states of node $\phi (x_{i} )$, an edge $e(i,j)$ between $i,j\in V$ represents a situation when two states $q,q'\in {\rm {\mathcal I}}$ are interfering with respect to a given state in $f\in {\rm {\mathcal Z}}_{i} $, thus vertices $i,j$ require different colors. 

Let 
\begin{equation} \label{12)} 
{\rm {\mathcal Z}}^{{\rm *}} ={\rm {\mathcal Z}}_{0} \bigcup {\rm {\mathcal Z}}_{1} ,\ldots ,\bigcup {\rm {\mathcal Z}}_{i} , i=0,\ldots ,|V|-1,                             
\end{equation} 
and a set 
\begin{equation} \label{13)} 
{\rm {\mathcal I}}^{{\rm *}} ={\rm {\mathcal I}}_{0} \bigcup {\rm {\mathcal I}}_{1} ,\ldots ,\bigcup {\rm {\mathcal I}}_{i} , i=0,\ldots ,|V|-1,                             
\end{equation} 
of all interfering entangled states, where 
\begin{equation} \label{14)} 
|{\rm {\mathcal I}}^{{\rm *}} |=k^{{\rm *}} .                                                   
\end{equation} 
Specifically, the problem for a given $N=(V,{\rm {\mathcal S}}^{{\rm *}} )$ is therefore that the conflicting vertices of ${\rm {\mathcal Z}}^{{\rm *}} $ have to be assigned to different states from a corresponding set ${\rm {\mathcal I}}^{{\rm *}} $.

Without loss of generality, since the sets ${\rm {\mathcal Z}}_{i} $ depends only on the determination of the link set ${\rm {\mathcal S}}^{{\rm *}} $, using a $k$-dimensional $n$-size base-graph $G^{k} $, the entanglement assignation problem for a given $U_{k} $ can be solved in at most ${\rm {\mathcal O}}(\log n)^{2} $ steps by our algorithm. As follows, if it is feasible to color vertices from $G$ using $k^{{\rm *}} $ different colors, then any valid coloring proper the assignment of the $k^{{\rm *}} $ different states from ${\rm {\mathcal Z}}^{{\rm *}} $ for the interfering states of ${\rm {\mathcal I}}^{{\rm *}} $. 

\begin{center}
\begin{figure*}[!h]
%\vspace{-0.5cm}
\begin{center}
\includegraphics[angle = 0,width=0.8\linewidth]{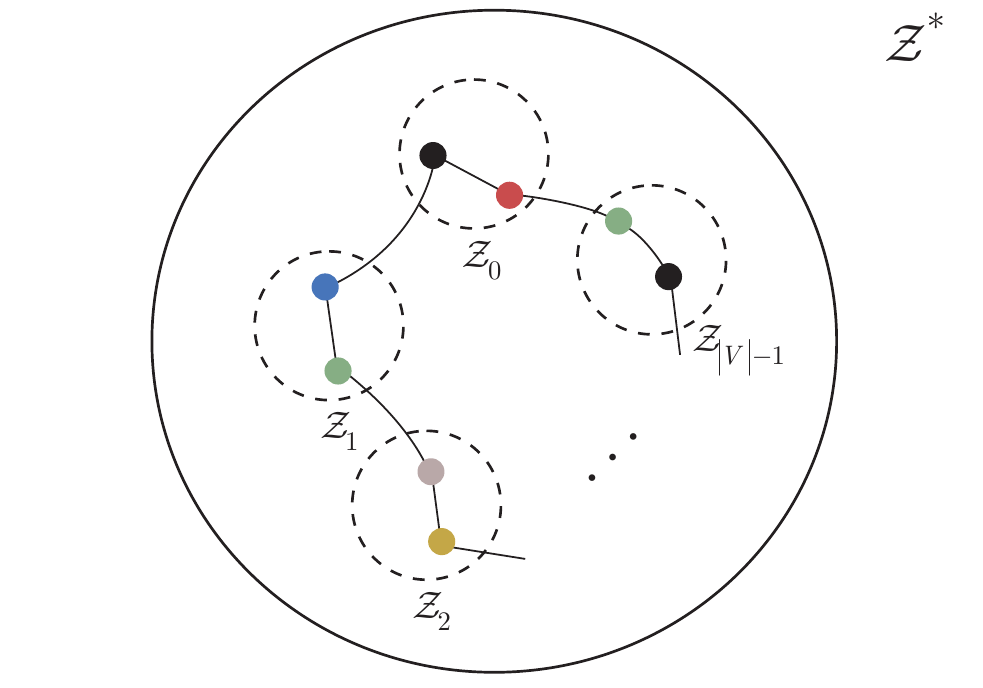}
\caption{A graph of conflicts of set ${\rm {\mathcal Z}}^{{\rm *}} ={\rm {\mathcal Z}}_{0} \bigcup {\rm {\mathcal Z}}_{1} ,\ldots ,\bigcup {\rm {\mathcal Z}}_{i} $ at $|{\rm {\mathcal I}}^{{\rm *}} |=k^{{\rm *}} $, where ${\rm {\mathcal I}}^{{\rm *}} ={\rm {\mathcal I}}_{0} \bigcup {\rm {\mathcal I}}_{1} ,\ldots ,\bigcup {\rm {\mathcal I}}_{i} $, $i=0,\ldots ,|V|-1$, of an entangled overlay quantum network $N=(V,{\rm {\mathcal S}}^{{\rm *}} )$. The vertices connected by edge represent interfering states. Since there are $k^{{\rm *}} $ interfering states in ${\rm {\mathcal I}}^{{\rm *}} $, the vertices of $G$ have to be colored by exactly $k^{{\rm *}} $ different colors such that connected vertices have to receive different colors.} 
 \label{fig3}
 \end{center}
\end{figure*}
\end{center}

The graph coloring problem \cite{ref15} for an entangled overlay quantum network $N=(V,{\rm {\mathcal S}}^{{\rm *}} )$ at ${\rm {\mathcal Z}}^{{\rm *}} $ and $|{\rm {\mathcal I}}^{{\rm *}} |=k^{{\rm *}} $ is depicted in \fref{fig3}.

In particular, these prove that the known NP-complete problem of vertex-coloring of a graph of conflicts can be transformed in polynomial time to the entanglement assignment problem, thus it polynomially reduces to our problem.

These facts prove that the entanglement assignment problem is NP-complete; therefore, to provide a time-efficient solution, an optimized approach is required. 

In our case, the time complexity is determined only by ${\rm {\mathcal A}}$, as the problem is to find a shortest path between a neighbor of the intermediate node, which is not the target node, and the target node. As follows, the overall complexity to solve \eqref{ZEqnNum542690} is bounded from above by ${\rm {\mathcal O}}(\log n)^{2} $ \cite{ref8}. If the intermediate node has no other neighbor besides the target node, then no other path exists between the intermediate node and the target node, requiring the establishment of a new entangled state between the intermediate node and the target node.

\section{Conclusions}
\label{sec4}
We defined a dynamic topology adaption algorithm for the quantum Internet. The method determines a set of entangled links to establish a new shortest path between the nodes. We utilized a base-graph construction for the efficient determination of the shortest paths required by the parallel demands of a multiuser network scenario. We studied the problem of entanglement assignment and defined an optimization procedure. The results can be applied directly for routing optimization problems in entangled quantum networks and allow for the provision of reliable quantum communications in the presence of time-varying stability quantum links of the quantum Internet.

\section*{Acknowledgements}
This work was partially supported by the National Research Development and Innovation Office of Hungary (Project No. 2017-1.2.1-NKP-2017-00001), by the Hungarian Scientific Research Fund - OTKA K-112125 and in part by the BME Artificial Intelligence FIKP grant of EMMI (BME FIKP-MI/SC).

\newpage
%\onecolumn
\appendix
\setcounter{table}{0}
\setcounter{figure}{0}
\setcounter{equation}{0}
\setcounter{algocf}{0}
\renewcommand{\thetable}{\Alph{section}.\arabic{table}}
\renewcommand{\thefigure}{\Alph{section}.\arabic{figure}}
\renewcommand{\theequation}{\Alph{section}.\arabic{equation}}
\renewcommand{\thealgocf}{\Alph{section}.\arabic{algocf}}

\setlength{\arrayrulewidth}{0.1mm}
\setlength{\tabcolsep}{5pt}
\renewcommand{\arraystretch}{1.5}
\section{Appendix}

\subsection{Notations}
The notations of the manuscript are summarized in \tref{tab2}.
\begin{center}
\begin{longtable}{||l|p{4.5in}||}
\caption{Summary of notations.}
\label{tab2}
\endfirsthead
\endhead
\hline
\textit{Notation} & \textit{Description} \\ \hline
L1 & Manhattan distance (L1 metric). \\ \hline 
$l$  & Level of entanglement.  \\ \hline 
${\rm L}_{l} $ & An $l$-level entangled link. For an ${\rm L}_{l} $ link, the hop-distance is $2^{l-1} $. \\ \hline 
$d(x,y)_{{\rm L}_{l} } $ & Hop-distance of an $l$-level entangled link between nodes $x$ and $y$.  \\ \hline 
$F$ & Fidelity of entanglement.  \\ \hline 
${\rm L}_{1} $ & ${\rm L}_{1} $-level (direct) entanglement,  $d(x,y)_{{\rm L}_{1} } =2^{0} =1$. \\ \hline 
${\rm L}_{2} $ & ${\rm L}_{2} $-level entanglement, $d(x,y)_{{\rm L}_{2} } =2^{1} =2$. \\ \hline 
${\rm L}_{3} $ & ${\rm L}_{3} $-level entanglement, $d(x,y)_{{\rm L}_{3} } =2^{2} =4$. \\ \hline 
$E(x,y)$ & An edge between quantum nodes $x$ and $y$, refers to an ${\rm L}_{l} $-level entangled link. \\ \hline 
${{\Pr }_{{{\text{L}}_{l}}}}( E( x,y ) )$ & Probability of existence of $E(x,y)$, $0<{{\Pr }_{{{\text{L}}_{l}}}}( E( x,y ) )\le 1$. \\ \hline 
$N$ & Overlay quantum network, $N=(V,E)$, where $V$ is the set of nodes, $E$ is the set of edges. \\ \hline 
$V$ & Set of nodes of $N$. \\ \hline 
$E$ & Set of edges of $N$. \\ \hline 
$G^{k} $ & An $n$-size, $k$-dimensional base-graph. \\ \hline 
$n$ & Size of base-graph $G^{k} $. \\ \hline 
$k$  & Dimension of base-graph $G^{k} $. \\ \hline 
$A$ & Transmitter node, $A\in V$. \\ \hline 
$B$ & Receiver node, $B\in V$. \\ \hline 
$R_{i} $ & A repeater node in $V$, $R_{i} \in V$. \\ \hline 
$E_{j} $ & Identifies an ${\rm L}_{l} $-level entanglement, $l=1,\ldots ,r$, between quantum nodes $x_{j} $ and $y_{j} $. \\ \hline 
$E=\{E_{j} \}$ & Let $E=\{E_{j} \}$, $j=1,\ldots ,m$ refer to a set of edges between the nodes of $V$. \\ \hline 
$\phi (x)$ & Position assigned to an overlay quantum network node $x\in V$ in a $k$-dimensional, $n$-sized finite square-lattice base-graph $G^{k} $. \\ \hline 
$\phi :V\to G^{k} $ & Mapping function that achieves the mapping from $V$ onto $G^{k} $. \\ \hline 
$d(\phi (x),\phi (y))$ & L1 distance between $\phi (x)$ and $\phi (y)$ in $G^{k} $. For   $\phi (x)=(j,k)$, $\phi (y)=(m,o)$ evaluated as\newline $d((j,k),(m,o))=|m-j|+|o-k|$. \\ \hline 
$p(\phi (x),\phi (y))$ & The probability that $\phi (x)$ and $\phi (y)$ are connected through an ${\rm L}_{l} $-level entanglement in $G^{k} $. \\ \hline 
$H_{n} $ & Normalizing term, defined as $H_{n} =\sum _{z}d(\phi (x),\phi (z)) $. \\ \hline 
$c_{\phi (x),\phi (y)} $ & Constant defined as\newline $c_{\phi (x),\phi (y)} ={{\Pr }_{{{\text{L}}_{l}}}}( E( x,y ) )-\frac{d(\phi (x),\phi (y))^{-k} }{H_{n} } ,$\newline where ${{\Pr }_{{{\text{L}}_{l}}}}( E( x,y ) )$ is the probability that nodes $x,y\in V$ are connected through an ${\rm L}_{l} $-level entanglement in the overlay quantum network $N$. \\ \hline 
${\rm {\mathcal R}}$ & Topology adaption algorithm in base-graph $G^{k} $. \\ \hline 
${\rm {\mathcal S}}^{{\rm *}} $ & Updated set of links, ${\rm {\mathcal S}}^{{\rm *}} =\{E_{i}^{{\rm *}} \}$, $i=0,\ldots ,h^{{\rm *}} -1$ of ${\rm L}_{l} $-level entangled links, $l=1,2,\ldots ,r$. Set of entangled links that have the highest probability in the network. \\ \hline 
$Q^{(F)} (\cdot )$ & Entanglement throughput. Number of $d$-dimensional maximally entangled states per second of a particular fidelity $F$. \\ \hline 
$\Pr _{{\rm L}_{l} }^{{\rm *}} $ & Threshold for a given  ${\rm L}_{l} $-level entangled link $E(x,y)$ from set ${\rm {\mathcal S}}^{{\rm *}} $ to determine the actual shortest path,  $\Pr _{{\rm L}_{l} } (x,y)\ge \Pr _{{\rm L}_{l} }^{{\rm *}} $. \\ \hline 
$D(\phi (A),\phi (B))$ & Diameter between source node $\phi (A)$ and target node $\phi (B)$. Refers to the maximum value of the shortest path (total number of edges on a path) between $\phi (A)$ and $\phi (B)$. \\ \hline 
${\rm {\mathcal L}}_{h} $ & Resource set, a set of $d$-dimensional maximally entangled states shared between $\phi (x_{i} ),\phi (x_{j} )$ through link $E_{h} $, \newline ${\rm {\mathcal L}}_{h} =\{{| \psi _{0}  \rangle} ,\ldots ,{| \psi _{g-1}  \rangle} \}$,\newline where $g$ is the number of entangled states established between nodes $\phi (x_{i} )$ and $\phi (x_{j} )$ through $E_{h} $. \\ \hline 
$\zeta (C)$ & Parameter subject to a minimization.  \\ \hline 
$U_{k} $ & User.  \\ \hline 
$K$ & Number of users. \\ \hline 
$\Pr _{{\rm L}_{l} } (E_{h}^{{\rm *}} )$ & Probability of existence of an ${\rm L}_{l} $-level entangled link $E_{h}^{{\rm *}} $ in the overlay network $N$. \\ \hline 
$p_{h}^{{\rm *}} $ & Probability of the ${\rm L}_{l} $-level entangled link $E_{h}^{{\rm *}} $ in the base-graph $G^{k} $. \\ \hline 
$C_{k,h}^{f} $ & Variable, equals 1 if the $f\in {\rm {\mathcal L}}_{h} $ maximally entangled resource state ${| \psi _{f}  \rangle} $ is assigned to the $k$-th user $U_{k} $ through link $E_{h}^{{\rm *}} $, 0 otherwise. \\ \hline 
$Q^{(F)} (U_{k} )$ & A demand of user $U_{k} $ with respect to the number of $d$-dimensional maximally entangled states per second of a particular fidelity F through link $E_{h}^{{\rm *}} $. \\ \hline 
$\phi (x_{t} )_{U_{k} } \in G^{k} $ & A node, associated to a given user $k\in K$ such that $E_{h}^{{\rm *}} (\phi (x_{t} )_{U_{k} } ,\phi (x_{j} )_{U_{k} } )$ is a link incident out of $\phi (x_{t} )_{U_{k} } \in G^{k} $, while $E_{h}^{{\rm *}} (\phi (x_{i} )_{U_{k} } ,\phi (x_{t} )_{U_{k} } )$ is a link incident onto node $\phi (x_{t} )_{U_{k} } \in G^{k} $ in the base-graph. \\ \hline 
$\Delta (C_{k,h}^{f} )$ & Parameter for flow conservation rules (from Kirchhoff's law). \\ \hline 
${\rm {\mathcal W}}_{h,j} $ & Parameter for flow conservation rules (from Kirchhoff's law). \\ \hline 
${\rm {\mathcal W}}_{h,i} $ & Parameter for flow conservation rules (from Kirchhoff's law). \\ \hline 
$\phi (A)_{U_{k} } $ & A source node of $U_{k} $. \\ \hline 
$\phi (B)_{U_{k} } $ & A target node of $U_{k} $. \\ \hline 
$\phi (s_{i} )$ & Source node of user $U_{i} $. \\ \hline 
$\phi (s_{j} )$ & Source node of user $U_{j} $. \\ \hline 
$\phi (x_{i} )$ & An intermediate node. \\ \hline 
$f$ & Resource state, an entangled state from the ${\rm {\mathcal L}}_{h} $ resource set. \\ \hline 
$q$ & A demand from a source node $\phi (s_{i} )$. \\ \hline 
$q'$ & An interfering demand from a source node $\phi (s_{j} )$. \\ \hline 
${\rm {\mathcal I}}$ & Interference set, a set of interfering demands  with respect to resource state $f\in {\rm {\mathcal L}}_{h} $. \\ \hline 
$K_{k,h}^{f} $ & A variable, equal to 1 if $q\in {\rm {\mathcal I}}$ is assigned to resource state $f\in {\rm {\mathcal L}}_{h} $, 0 otherwise. \\ \hline 
$k$ & Cardinality of interference set ${\rm {\mathcal I}}$, $|{\rm {\mathcal I}}|=k$. \\ \hline 
$G=(V,E)$ & A graph of conflicts, where $V$ is a set of vertices, while $E$ is a set of edges, $|E|=|{\rm {\mathcal S}}^{{\rm *}} |$. \\ \hline 
$e(i,j)\in E$ & An edge in a graph of conflicts $G=(V,E)$, the vertices $i,j$ are conflicting vertices. \\ \hline 
${\rm {\mathcal Z}}_{i} $ & A set of entangled states of node $\phi (x_{i} )$.  \\ \hline 
${\rm {\mathcal Z}}^{{\rm *}} $ & A set of vertices of in a graph of conflicts $G$, ${\rm {\mathcal Z}}^{{\rm *}} ={\rm {\mathcal Z}}_{0} \bigcup {\rm {\mathcal Z}}_{1} ,\ldots ,\bigcup {\rm {\mathcal Z}}_{i} $, $i=0,\ldots ,|V|-1$.      \\ \hline 
${\rm {\mathcal I}}^{{\rm *}} $ & Set of interfering entangled states, ${\rm {\mathcal I}}^{{\rm *}} ={\rm {\mathcal I}}_{0} \bigcup {\rm {\mathcal I}}_{1} ,\ldots ,\bigcup {\rm {\mathcal I}}_{i} $, $i=0,\ldots ,|V|-1$.   \\ \hline 
$k^{{\rm *}} $ & Variable, identifies the number of interfering states in set ${\rm {\mathcal I}}^{{\rm *}} $.  \\ \hline 
\end{longtable}
\end{center}

\begin{thebibliography}{10}
\bibitem{ref1} Van Meter, R. \textit{Quantum Networking}. ISBN 1118648927, 9781118648926, John Wiley and Sons Ltd (2014).

\bibitem{ref2} Lloyd, S., Shapiro, J.H., Wong, F.N.C., Kumar, P., Shahriar, S.M. and Yuen, H.P. Infrastructure for the quantum Internet. \textit{ACM SIGCOMM} \textit{Computer} \textit{Communication Review}, 34, 9--20 (2004).

\bibitem{ref3} Pirandola, S. Capacities of repeater-assisted quantum communications, \textit{arXiv:1601.00966} (2016).

\bibitem{ref4} Kimble, H.J. The quantum Internet. \textit{Nature}, 453:1023--1030 (2008).

\bibitem{ref5} Gyongyosi, L., Imre, S. and Nguyen, H.V. A Survey on Quantum Channel Capacities, \textit{IEEE Communications Surveys and Tutorials}, doi: 10.1109/COMST.2017.2786748 (2018).

\bibitem{ref6} Munro, W.J., Stephens, A.M., Devitt, S.J., Harrison, K.A. and Nemoto, K. Quantum communication without the necessity of quantum memories, \textit{Nature Photonics} 6, 777- 781 (2012). 

\bibitem{ref7} Kok, P., Munro, W.J., Nemoto, K., Ralph, T.C., Dowling, J.P. and Milburn, G.J., Linear optical quantum computing with photonic qubits, \textit{Rev. Mod. Phys}. 79, 135-174 (2007).

\bibitem{ref8} Gyongyosi, L. and Imre, S. Decentralized Base-Graph Routing for the Quantum Internet, \textit{Physical Review A}, American Physical Society, DOI: 10.1103/PhysRevA.98.022310, https://link.aps.org/doi/10.1103/PhysRevA.98.022310, 2018.

\bibitem{ref9} Gyongyosi, L. and Imre, S. Dynamic topology resilience for quantum networks, \textit{Proc. SPIE 10547}, Advances in Photonics of Quantum Computing, Memory, and Communication XI, 105470Z (22 February 2018); doi: 10.1117/12.2288707.

\bibitem{ref10} Pirandola, S., Laurenza, R., Ottaviani, C. and Banchi, L. Fundamental limits of repeaterless quantum communications, \textit{Nature Communications}, 15043, doi:10.1038/ncomms15043 (2017).

\bibitem{ref11} Pirandola, S., Braunstein, S.L., Laurenza, R., Ottaviani, C., Cope, T.P.W., Spedalieri, G. and Banchi, L. Theory of channel simulation and bounds for private communication, \textit{Quantum Sci. Technol}. 3, 035009 (2018).

\bibitem{ref12} Laurenza, R. and Pirandola, S. General bounds for sender-receiver capacities in multipoint quantum communications, \textit{Phys. Rev. A} 96, 032318 (2017).

\bibitem{ref13} Kleinberg, J. The Small-World Phenomenon: An Algorithmic Perspective, \textit{Proceedings of the 32nd Annual ACM Symposium on Theory of Computing} (STOC'00), (2000).

\bibitem{ref14} Franceschetti, M. and Meester, R. \textit{Random Networks for Communication}, Cambridge University Press (2008).

\bibitem{ref15} Rak, J. \textit{Resilient Routing in Communication Networks}, Springer (2015).

\bibitem{ref16} Gyongyosi, L. and Imre, S. Multilayer Optimization for the Quantum Internet, \textit{Scientific Reports}, Nature, DOI:10.1038/s41598-018-30957-x, 2018.

\bibitem{ref17} Gyongyosi, L. and Imre, S. Entanglement Availability Differentiation Service for the Quantum Internet, \textit{Scientific Reports}, Nature, (DOI:10.1038/s41598-018-28801-3), https://www.nature.com/articles/s41598-018-28801-3, 2018.

\bibitem{ref18} Gyongyosi, L. and Imre, S. Entanglement-Gradient Routing for Quantum Networks, \textit{Scientific Reports}, Nature, (DOI:10.1038/s41598-017-14394-w), https://www.nature.com/articles/s41598-017-14394-w, 2017.

\bibitem{ref19} Imre, S. and Gyongyosi, L. \textit{Advanced Quantum Communications - An Engineering Approach}. New Jersey, Wiley-IEEE Press (2013).

\bibitem{ref20} Caleffi, M. End-to-End Entanglement Rate: Toward a addressStreetQuantum Route Metric, 2017 \textit{IEEE Globecom}, DOI: 10.1109/GLOCOMW.2017.8269080, (2018). 

\bibitem{ref21} Van Meter, R., Satoh, T., Ladd, T.D., Munro, W.J. and Nemoto, K. Path selection for quantum repeater networks, \textit{Networking Science}, Volume 3, Issue 1--4, pp 82--95, (2013).

\bibitem{ref22} Caleffi, M. Optimal Routing for Quantum Networks, \textit{IEEE Access}, Vol 5, DOI: 10.1109/ACCESS.2017.2763325 (2017).

\bibitem{ref23} Caleffi, M., Cacciapuoti, A.S. and Bianchi, G. Quantum Internet: from Communication to Distributed Computing, \textit{aXiv:1805.04360} (2018).

\bibitem{ref24} Castelvecchi, D. The quantum internet has arrived, \textit{Nature}, News and Comment, https://www.nature.com/articles/d41586-018-01835-3, (2018).

\bibitem{ref25} Petz, D. \textit{Quantum Information Theory and Quantum Statistics}, Springer-Verlag, Heidelberg, Hiv: 6. (2008).

\bibitem{ref26} Bacsardi, L. On the Way to Quantum-Based Satellite Communication, \textit{IEEE Comm. Mag.} 51:(08) pp. 50-55. (2013).

\bibitem{ref27} Biamonte, J. et al. Quantum Machine Learning. \textit{Nature}, 549, 195-202 (2017). 

\bibitem{ref28} Lloyd, S., Mohseni, M. and Rebentrost, P. Quantum algorithms for supervised and unsupervised machine learning. \textit{arXiv:1307.0411} (2013).

\bibitem{ref29} Lloyd, S., Mohseni, M. and Rebentrost, P. Quantum principal component analysis. \textit{Nature Physics}, 10, 631 (2014).

\bibitem{ref30} Lloyd, S. Capacity of the noisy quantum channel. \textit{Physical Rev. A}, 55:1613--1622 (1997).

\bibitem{ref31} Lloyd, S. The Universe as Quantum Computer, \textit{A Computable Universe: Understanding and exploring Nature as computation}, Zenil, H. ed., World Scientific, Singapore, \textit{arXiv:1312.4455v1} (2013).

\bibitem{ref32} Shor, P.W. Scheme for reducing decoherence in quantum computer memory. \textit{Phys. Rev. A}, 52, R2493-R2496 (1995).

\bibitem{ref33} Chou, C., Laurat, J., Deng, H., Choi, K. S., de Riedmatten, H., Felinto, D. and Kimble, H.J. Functional quantum nodes for entanglement distribution over scalable quantum networks. \textit{Science}, 316(5829):1316--1320 (2007).

\bibitem{ref34} Muralidharan, S., Kim, J., Lutkenhaus, N., Lukin, M.D. and Jiang. L. Ultrafast and Fault-Tolerant Quantum Communication across Long Distances, \textit{Phys. Rev. Lett}. 112, 250501 (2014).

\bibitem{ref35} Van Meter, R., Ladd, T. D., Munro, W.J. and Nemoto, K. System Design for a Long-Line Quantum Repeater, \textit{IEEE/ACM Transactions on Networking} 17(3), 1002-1013, (2009).

\bibitem{ref36} Gisin, N. and Thew, R. Quantum Communication. \textit{Nature Photon.} 1, 165-171 (2007).

\bibitem{ref37} Yuan, Z., Chen, Y., Zhao, B., Chen, S., Schmiedmayer, J. and Pan, J.W. \textit{Nature} 454, 1098-1101 (2008).

\bibitem{ref38} Kobayashi, H., Le Gall, F., Nishimura, H. and Rotteler, M. General scheme for perfect quantum network coding with free classical, \textit{Communication Lecture Notes in Computer Science} (Automata, Languages and Programming SE-52 vol. 5555), Springer) pp 622-633 (2009).

\bibitem{ref39} Leung, D., Oppenheim, J. and Winter, A. \textit{IEEE Trans. Inf. Theory} 56, 3478-90. (2010).

\bibitem{ref40} Kobayashi, H., Le Gall, F., Nishimura, H. and Rotteler, M. Perfect quantum network communication protocol based on classical network coding, \textit{Proceedings of 2010 IEEE International Symposium on Information Theory} (ISIT) pp 2686-90. (2010).

\bibitem{ref41} Hayashi, M. Prior entanglement between senders enables perfect quantum network coding with modification, \textit{Physical Review A}, Vol.76, 040301(R) (2007).

\bibitem{ref42} Hayashi, M., Iwama, K., Nishimura, H., Raymond, R. and Yamashita, S, Quantum network coding, \textit{Lecture Notes in Computer Science} (STACS 2007 SE52 vol. 4393) ed Thomas, W. and Weil, P. (Berlin Heidelberg: Springer) (2007).

\bibitem{ref43} Chen, L. and Hayashi, M. Multicopy and stochastic transformation of multipartite pure states, \textit{Physical Review A}, Vol.83, No.2, 022331, (2011).

\bibitem{ref44} Schoute, E., Mancinska, L., Islam, T., Kerenidis, I. and Wehner, S. Shortcuts to quantum network routing, \textit{arXiv:1610.05238} (2016).

\bibitem{ref45} Lloyd, S. and Weedbrook, C. Quantum generative adversarial learning. \textit{Phys. Rev. Lett}., 121, arXiv:1804.09139 (2018).

\end{thebibliography}
\end{document}